\documentclass[11pt]{aipproc}
\usepackage{epsfig}

\newcommand{\psb}{\bar{\psi}}

\newcommand{\ben}{\begin{enumerate}}
\newcommand{\een}{\end{enumerate}}
\newcommand{\bit}{\begin{itemize}}
\newcommand{\eit}{\end{itemize}}
\newcommand{\bc}{\begin{center}}
\newcommand{\ec}{\end{center}}
\newcommand{\bq}{\begin{equation}}
\newcommand{\eq}{\end{equation}}
\newcommand{\bqa}{\begin{eqnarray}}
\newcommand{\eqa}{\end{eqnarray}}

\newcommand{\nn}{\nonumber}
\newcommand{\nl}{\nonumber\\}

\newcommand{\plaat}[3]{\raisebox{#3pt}{
\epsfig{figure=./#1.eps,
width=#2cm}}}

\newcommand{\plaata}[4]{\raisebox{#2pt}{
\epsfig{figure=./#1.eps,
width=#3cm,height=#4cm}}}

\def\demo{$\Delta\eta\mu \acute{o} \kappa \varrho \iota \tau o \varsigma$}

\begin{document}

\title{ {\tt HELAC-PHEGAS}: automatic computation\\
of helicity amplitudes and cross sections}
\author{ Aggeliki Kanaki and Costas G.~Papadopoulos\thanks{%
Invited talk at ACAT2000, October 16-20, 2000, Fermilab, IL, USA.}}
\affiliation{Institute of Nuclear Physics, NCSR \demo, 15310 Athens, Greece}

\begin{abstract}
{\tt HELAC-PHEGAS} is a {\tt FORTRAN} based package that is able to compute
automatically and efficiently tree-order 
helicity amplitudes and cross sections for 
arbitrary scattering processes
within the standard electroweak theory and QCD. 
The algorithm for the
amplitude computation, {\tt HELAC},
exploits the virtues of the Dyson-Schwinger equations. 
The phase-space generation algorithm,
{\tt PHEGAS}, constructs all possible kinematical mappings dictated
by the amplitude under consideration. Combined with  
mutichannel self-optimized Monte Carlo integration 
it results to efficient cross section evaluation.
\end{abstract}

\maketitle

\section{Introduction} 

The need for efficient algorithms to calculate 
helicity amplitudes and
cross sections for any process, in an automatic way, 
has been well recognized long time ago.
Up to now, algorithms that efficiently combine helicity amplitude computation 
and phase-space integration have been prooven successful 
for specific processes, 
like for instance four-fermion~\cite{four-fermion} 
production in $e^+e^-$ collisions. On the other hand 
general-purpose computational packages like
{\tt CompHEP}~\cite{comphep} and {\tt GRACE}~\cite{grace}
do not provide automatic efficient phase-space integration algorithms. 
Moreover the vast majority of the automatized helicity amplitude 
computationial algorithms, like for instance {\tt MadGraph}~\cite{madg},
have been based on the Feynman graph representation of the amplitude which
severely restricts their ability to deal with multiparticle scattering 
processes.

In this article we report on some developments that have lead to the
construction of two programs, {\tt HELAC}~\cite{helac} and 
{\tt PHEGAS}~\cite{phegas}, that
allow for an efficient and automatic evaluation of cross sections
for arbitrary scattering processes.  

\section{\tt HELAC}

The traditional representation of the scattering amplitude in terms of
Feynman graphs results to a computational cost that grows like the
number of those graphs, therefore as $n!$ where $n$ is the
number of particles involved in the scattering process. 

An alternative\footnote{See also references~\cite{berends} and ~\cite{camo}.} 
to the Feynman graph representation is provided by the
Dyson-Schwinger approach. Dyson-Schwinger equations 
express recursively the $n$-point Green's functions
in terms of the $1-,2-,\ldots,(n-1)$-point functions.
For instance in QED these equations can be written as follows:

\[
\plaata{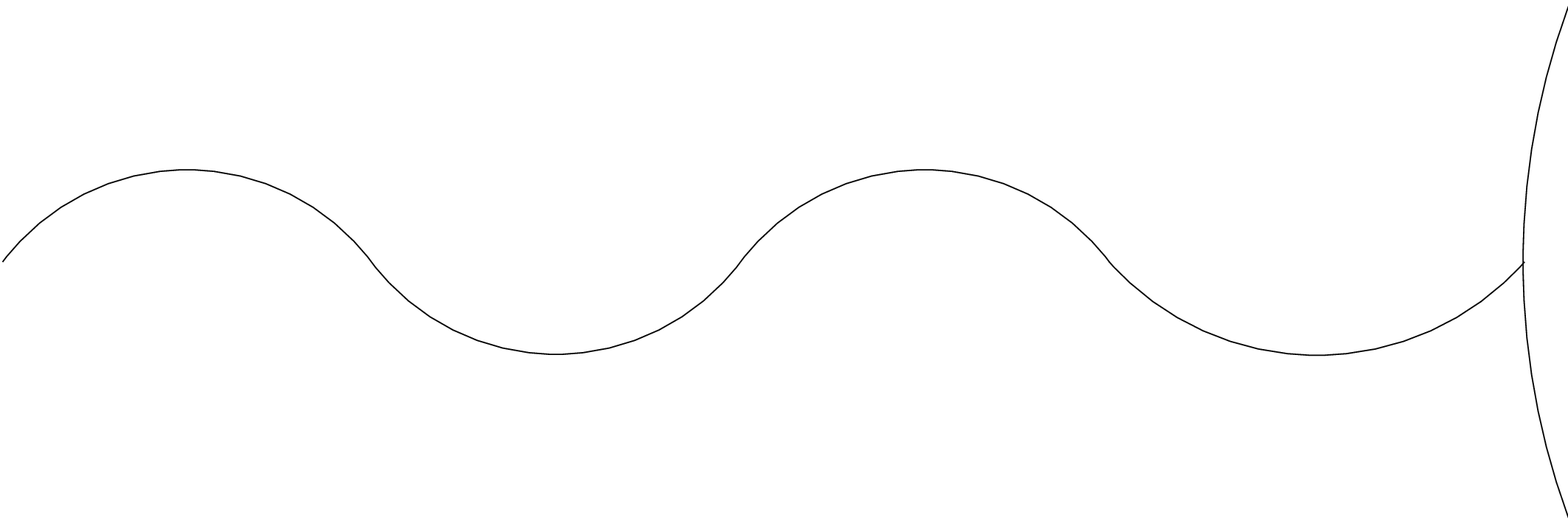}{-15}{1.8}{1}\;\;=\;\plaata{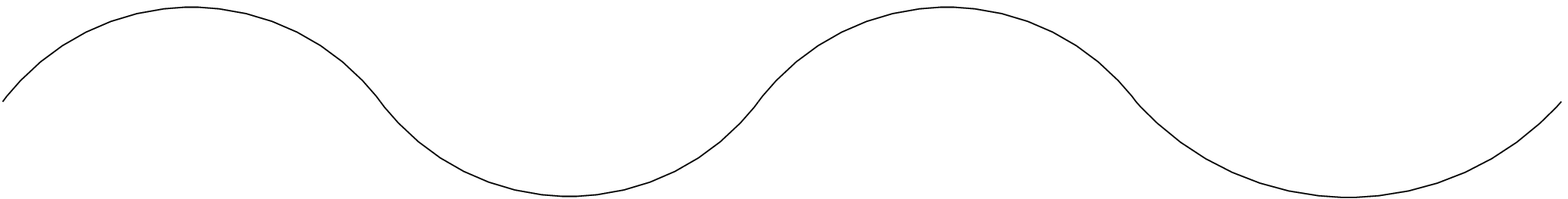}{0}{1.8}{0.2}\;
+\;\plaata{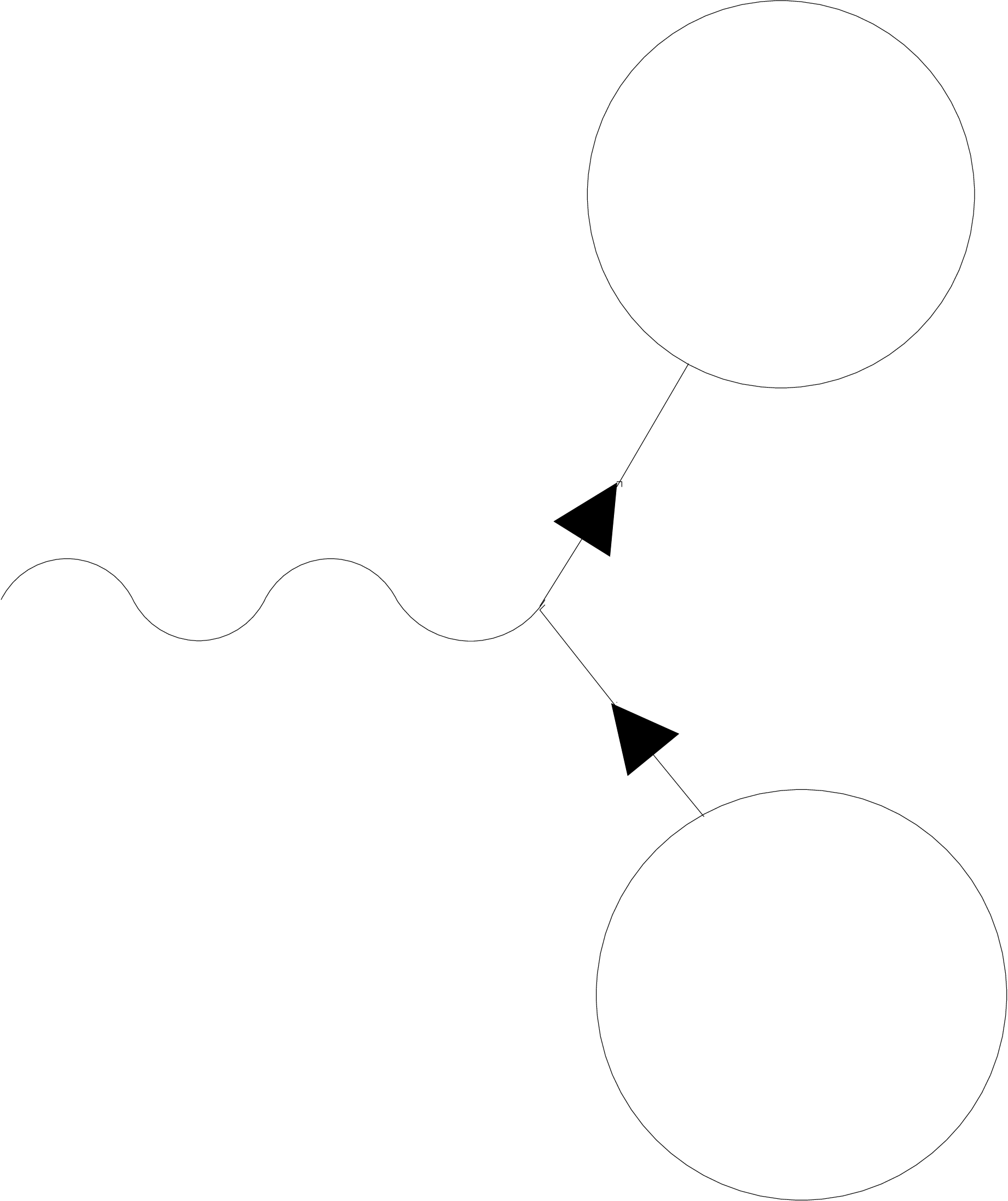}{-25}{1.8}{2}
\]
\bqa
&& b^\mu(P)=\sum_{i=1}^n
\delta_{P=p_i} b^\mu(p_i) 
\nn 
\\
&&+ \sum_{P=P_1+P_2}
(ig)\Pi^\mu_\nu\; \psb(P_2)\gamma^\nu\psi(P_1)
\epsilon(P_1,P_2) 
\nn
\eqa
where 
\[
 b_\mu(P) \;=\; \plaata{bos1}{-5}{0.8}{0.5} \;\;\;\;
   \psi(P)  \;=\; \plaata{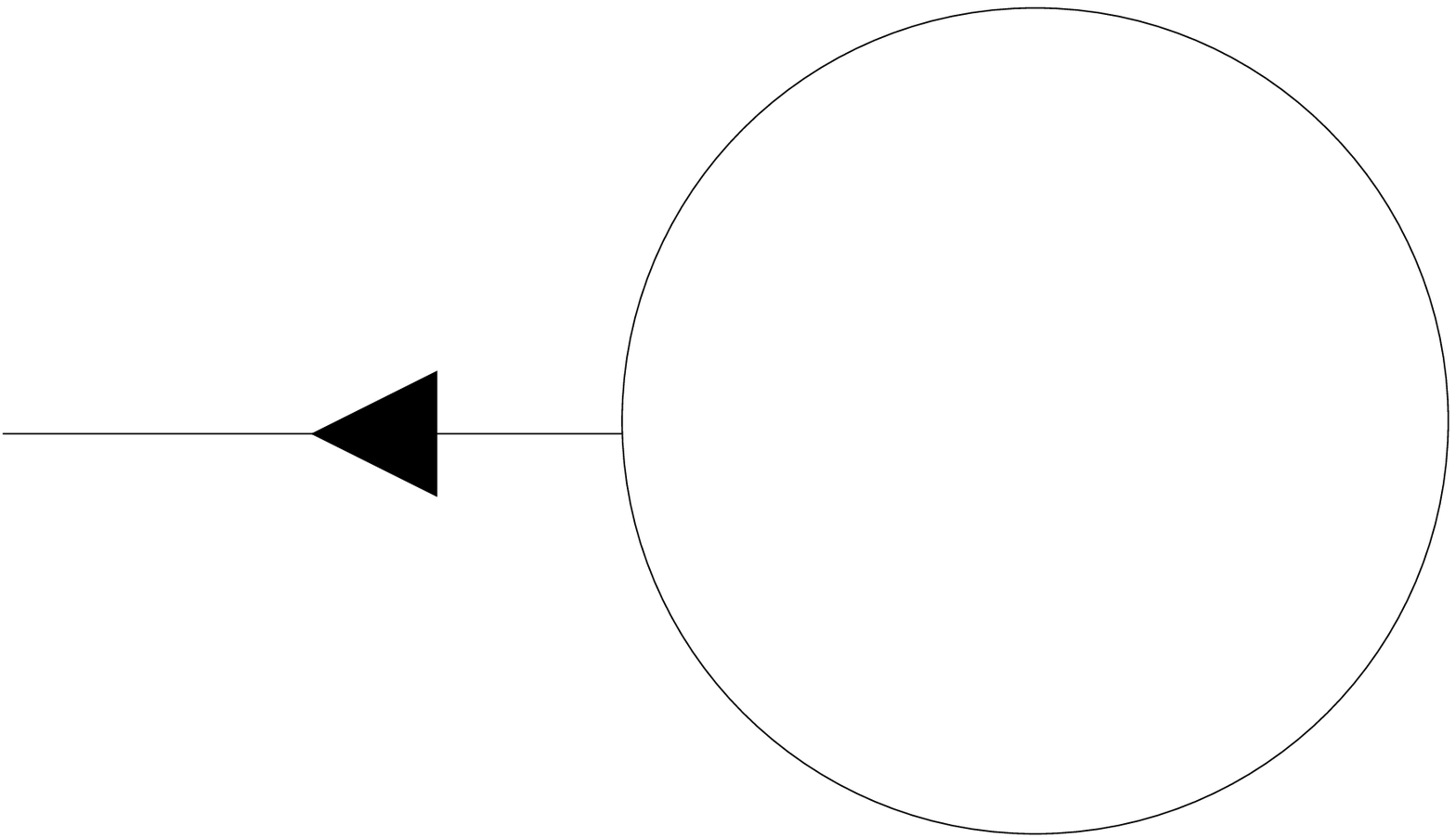}{-6}{0.8}{0.6}\; \;\;\;
   \psb(P)  \;=\; \plaata{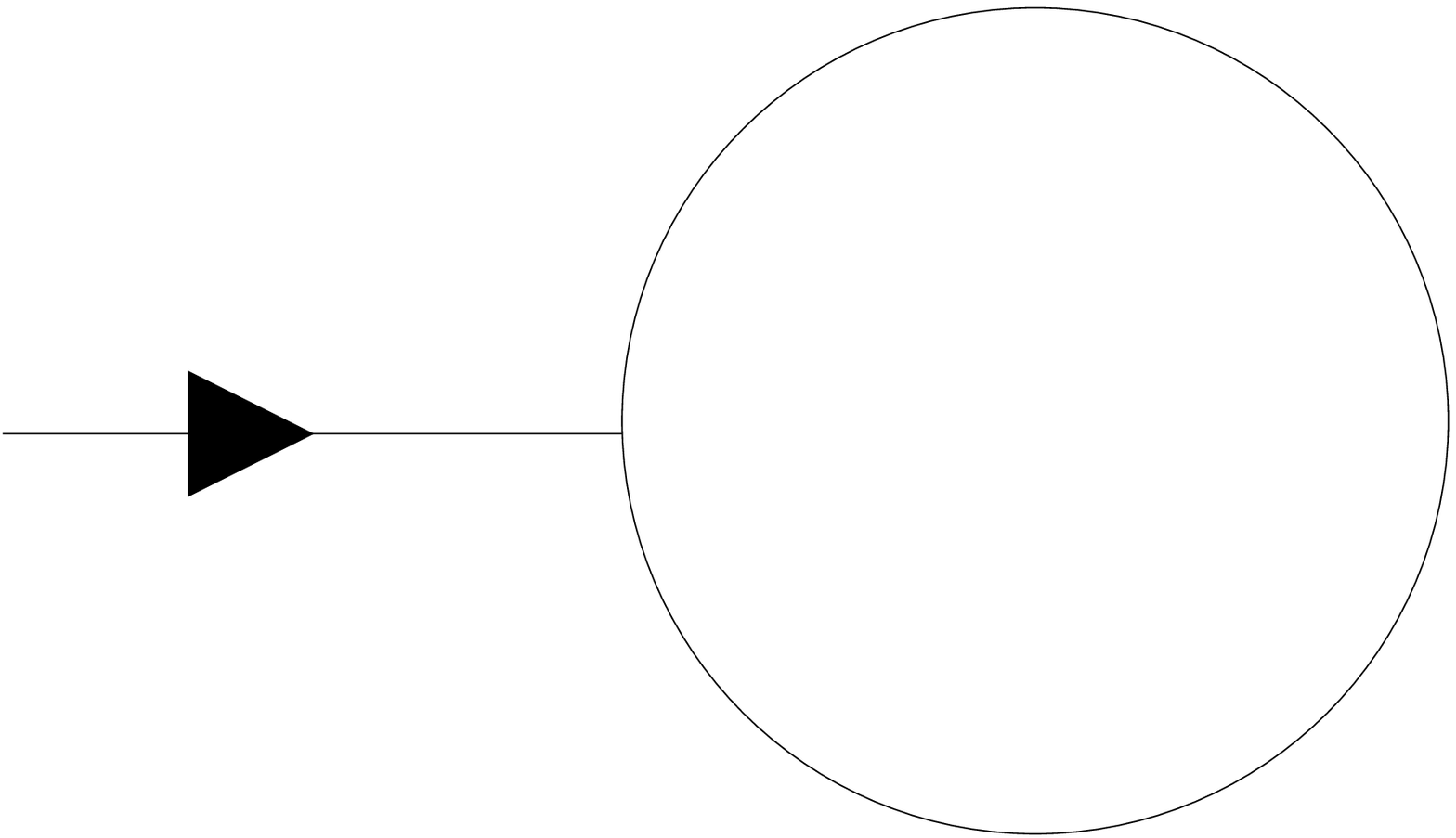}{-6}{0.8}{0.6} 
\]
describes a generic $n$-point Green's function 
with respectively one outgoing photon, fermion or anti\-fermion 
leg carrying momentum $P$. $\Pi_{\mu\nu}$ stands for the boson propagator
and $\epsilon$ takes into account the sign due to fermion antisymmetrization.

In order to actually solve these recursive equations it is convenient to use
a binary representation of the momenta involved~\cite{camo}.
For a process involving $n$ external particles, 
all momenta appearing in the computation, $P^\mu$,
\[
P^\mu=\sum_{i\in I}p_i^\mu
\]
where $I\subset \{ 1,\ldots,n\}$, 
can be assigned a binary vector $\vec{m}=(m_1,\ldots,m_n)$,
where its components take the values $0$ or $1$, in such a way that
\[
P^\mu=\sum_{i=1}^n m_i\;p_i^\mu\;.
\]
Moreover this binary vector can be uniquely represented by the
integer
\[
m=\sum_{i=1}^n 2^{i-1}m_i
\] 
and therefore all subamplitudes can be labeled accordingly,i.e.
\[ 
b_\mu(P)\to b_\mu(m)\;,\;1\le m \le 2^{n-1}.
\]
A very convenient ordering of integers in binary representation
relies on the notion of level $l$, defined simply as
\[
l=\sum_{i=1}^n m_i\;.
\]
As it is easily seen all external momenta are of level $1$, whereas
the total amplitude corresponds to the unique level $n$ integer
$2^{n-1}$. This ordering dictates the natural path of the
computation; starting with level-$1$ sub-amplitudes, we compute
the level-$2$ ones using the Dyson-Schwinger equations and so on
up to the level $n$ one which is the full amplitude. 
For the spinor wave functions as well as for the Dirac matrices, 
we have chosen the 4-dimenional
chiral representation which results to particurarly simple expressions. 
All electroweak vertices in both the Feynman and the Unitary gauge
have been included.

The computational cost of {\tt HELAC} grows like $\sim 3^n$, which essentially
counts the steps used to solve the recursive equations.
Obviously for large $n$ there is a tremendous saving of computational time,
compared to the $n!$ growth of the Feynman graph approach.

For QCD amplitudes colour representation and summation plays an important
role. 
Let $1\ldots n$ denote the colour labels of quarks and
$\sigma_i(1)\ldots\sigma_i(n)$ denote the colour labels of antiquarks, 
with $\sigma(i), i=1\ldots n!$ being a permutation of $\{1\ldots n\}$.
The colour factor is given obviously by
\[
{\cal C}_i=\delta_{1\sigma_i(1)}\delta_{2\sigma_i(2)}
\ldots\delta_{n\sigma_i(n)}
\]
Moreover the colour matrix, defined as 
\[ 
{\cal M}_{ij} = \sum_{\mbox{\scriptsize  colours}} C_i C_j^\dagger
\]
with the summation running over all colours, $1\ldots N_c$, has a very simple
representation
\[
{\cal M}_{ij} = N_c^{m(\sigma_i,\sigma_j)}
\]
where $m(\sigma_i,\sigma_j)-1$ counts how many elements of the permutations
$\sigma_i$ and $\sigma_j$ are common.
In order to extent this colour representation to QCD amplitudes we have just 
to consider gluons as being quark-antiquark pairs and assign to them
two colour labels $(i,\sigma_i)$. The colour factor and the colour matrix 
still has
exactly the same form. The only thing one has to consider is to rewrite
the known Feynman rules of QCD in a slight different way. 
It is worthwhile to note that exact colour summation is efficient as
far as the number of equivalent gluons is smaller than ${\cal O}(5-6)$.
For multicolour processes other approaches have to be considered~\cite{%
qcd,alphaqcd}.

The programme is also incorporating the possibility to use an extended
precision by exploiting the virtues of {\tt FORTRAN90}. The user can easily
switch to a quadruple precision or to an even  higher, user-defined 
precision by using the multi-precision library~\cite{mpcl} included
in {\tt HELAC}.
In this way, a straightforward computation of cross sections for processes
like $e^-e^+ \to e^-e^+e^-e^+$ without any cut is reliably performed~\cite{%
nextc}.

\section{\tt PHEGAS}

\begin{table}[t]
\begin{tabular}{lccr}
\hline
\parbox[t]{3cm}{\centerline{Final states}}
&
\parbox{3cm}{\centerline{Number of FG(DS)}}
&
\parbox{3cm}{\centerline{$\sqrt{s}$ (GeV) }}
&
\parbox{3cm}{\hfill Cross section (fb)}  
\\[6pt]
\hline
$e^-e^+\to u\ \bar{d}\ s\ \bar{c}\ \gamma$ 
&
90(74)
&
200
&
199.75 (16)
\\[6pt]
$e^-e^+\to  e^-\ \bar{\nu}_e\ \mu^+\ \nu_\mu\ \gamma $
&
108(100)
&
200
&
29.309 (25)
\\[6pt]
$e^-e^+\to \mu^-\ \bar{\nu}_\mu\ u\ \bar{d}\ \gamma\ \gamma$
&
587(210)
&
500
&
1.730 (58)
\\[6pt]
$e^-e^+\to \mu^-\ \bar{\nu}_\mu\ u\ \bar{d}\ c\ \bar{c} $
&
209(102)
&
500
&
0.1783 (20)
\\[6pt]
$e^-e^+\to \mu^-\ \bar{\nu}_\mu\ u\ \bar{d}\ c\ \bar{c}\ \gamma $
&
2142(339)
&
500
&
0.02451 (65)
\\[6pt]
$g\ g \to b\ \bar{b}\ b\ \bar{b}\ W^-\  W^+ $
&
960(380)
&
500
&
4.716(24)
\\[6pt]
\hline
\end{tabular}
\caption{Results for several processes using {\tt HELAC-PHEGAS}.
In the second column the number of Feynman graphs and in parenthesis
the number  
of steps required to solve the recursiveDyson-Schwinger equations are given.}
\label{tab1}
\end{table}

The study of multi-particle processes, like for instance four-fermion 
production in $e^+e^-$, requires efficient phase-space Monte Carlo generators.
The reason is that the squared amplitude, being a complicated function
of the kinemtaical variables, exhibits strong
variations in specific regions and/or directions of the phase space, 
lowering in a substantial
way the speed and the efficiency of the Monte Carlo integration. A well known
way out of this problem relies on  
algorithms characterized by two main ingredients:
\begin{enumerate}
\item The construction of appropriate mappings of the phase space 
parametrization in such a way that the main variation of the integrand
can be described by a set of almost uncorrelated  variables, and
\item A self-adaptation procedure that reshapes the generated phase-space
density in order to be as much as possible close to the integrand.
\end{enumerate}
In order to construct appropriate mappings we note that  
the integrand, i.e. the squared amplitude,  has a well-defined
representation in terms of Feynman diagrams. It is therefore natural
to associate to each Feynman diagram a phase-space mapping that 
parametrizes the leading variation coming from it. 
To be more specific the contribution 
of tree-order Feynman diagrams to the full amplitude can be factorized
in terms of propagators, vertex factors and external wave functions.
In general, the main source of variation comes from the propagator
factors and therefore our aim is to construct a mapping 
that expresses the phase-space
density in terms of the kinematical invariants that appear
in these propagator factors. 
Since in principle we need
as many mappings as Feynman diagrams for the process under consideration,
we have to appropriately combine them in order to produce
the global phase-space density. A simple and well studied solution
to this problem was suggested some time ago
in reference~\cite{multi}.
It should be 
mentioned however that other self-adapting approaches can be 
used as well~\cite{Ohl:1999jn}. 
It is important to note that although by using Feynman
graphs to construct phase-space mappings we face the original
$n!$ computational cost growth problem, the self-optimization cures to
a certain extent this by selecting only the few mappings that dominate
the phase-space density. 
For alternative approaches we refer to \cite{sarge}.

In order to describe the construction of the phase-space mappings,
let us consider a typical process in which two incoming particles
produce $n$ outgoing ones. The phase space, $d\Phi_n(P=q_1+q_2;p_1\ldots,p_n)$,
can be decomposed as follows
\bqa
&& 
d\Phi_{n}=\left(\prod_{i=1}^{m}\frac{dQ_i^2}{2\pi}\right)
d\Phi_{m}(P;Q_1,\ldots,Q_m)
\nl 
&&
d\Phi_{n_1}(Q_1;r_1,r_2,\ldots,r_{n_1})
\ldots
 d\Phi_{n_m}(Q_m;s_1,s_2,\ldots,s_{n_m})
\nn
\eqa
where the subsets $\{r_1,r_2,\ldots,r_{n_1}\}$ 
up to  $\{s_1,s_2,\ldots,s_{n_m}\}$ 
represent
an arbitrary partition of $\{p_1,p_2,\ldots,p_n\}$.
The above equation can be generalized recursively resulting 
in an arbitrary decomposition of $d\Phi_{n}$.
Feynman graphs can be seen as a realization
of such a decomposition, this latter being identified with a 
sequence of vertices of the graph. 
There two possible cases for $2\to n$ scattering.
First, all outgoing momenta involved in the vertex are time-like, 
\[
\plaat{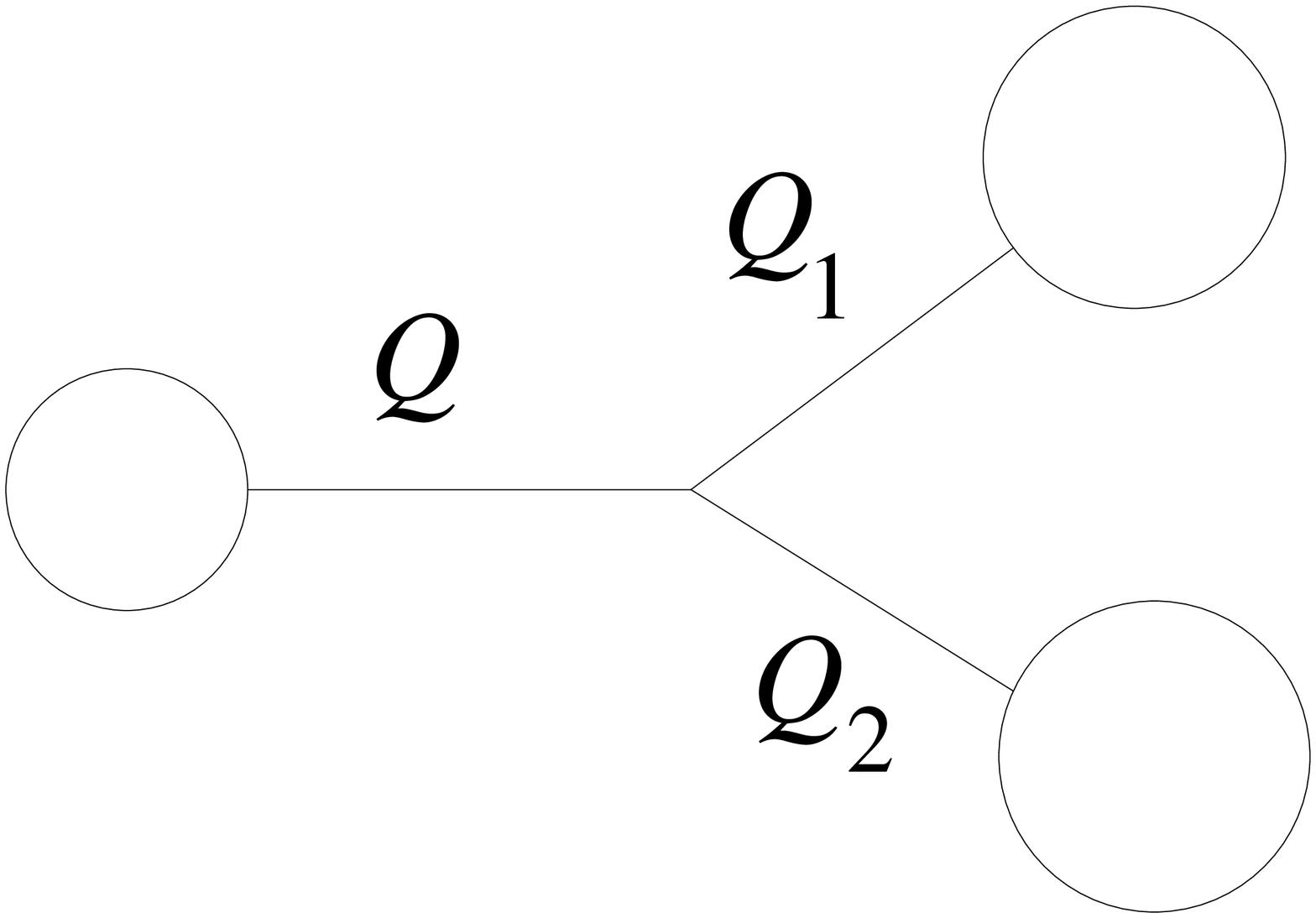}{3}{-24}
\]
\bqa
 d\Phi_n &=& \ldots \frac{d Q_1^2}{2 \pi}\; \frac{d Q_2^2}{2 \pi}\;
d\Phi_2(Q \to Q_1,Q_2) \ldots \nl
&=& \ldots \frac{d Q_1^2}{2 \pi}\; \frac{d Q_2^2}{2 \pi}\;
d \cos\theta \;d \phi\;
 \frac{\lambda^{1/2}(Q^2,Q_1^2,Q_2^2)}{32 \pi^2\; Q^2} \ldots
\nn
\eqa 
with
$\lambda(x,y,z)=x^2+y^2+z^2-2 xy-2 x z -2 yz$,
and second when one  of them is space-like,
\[
\plaat{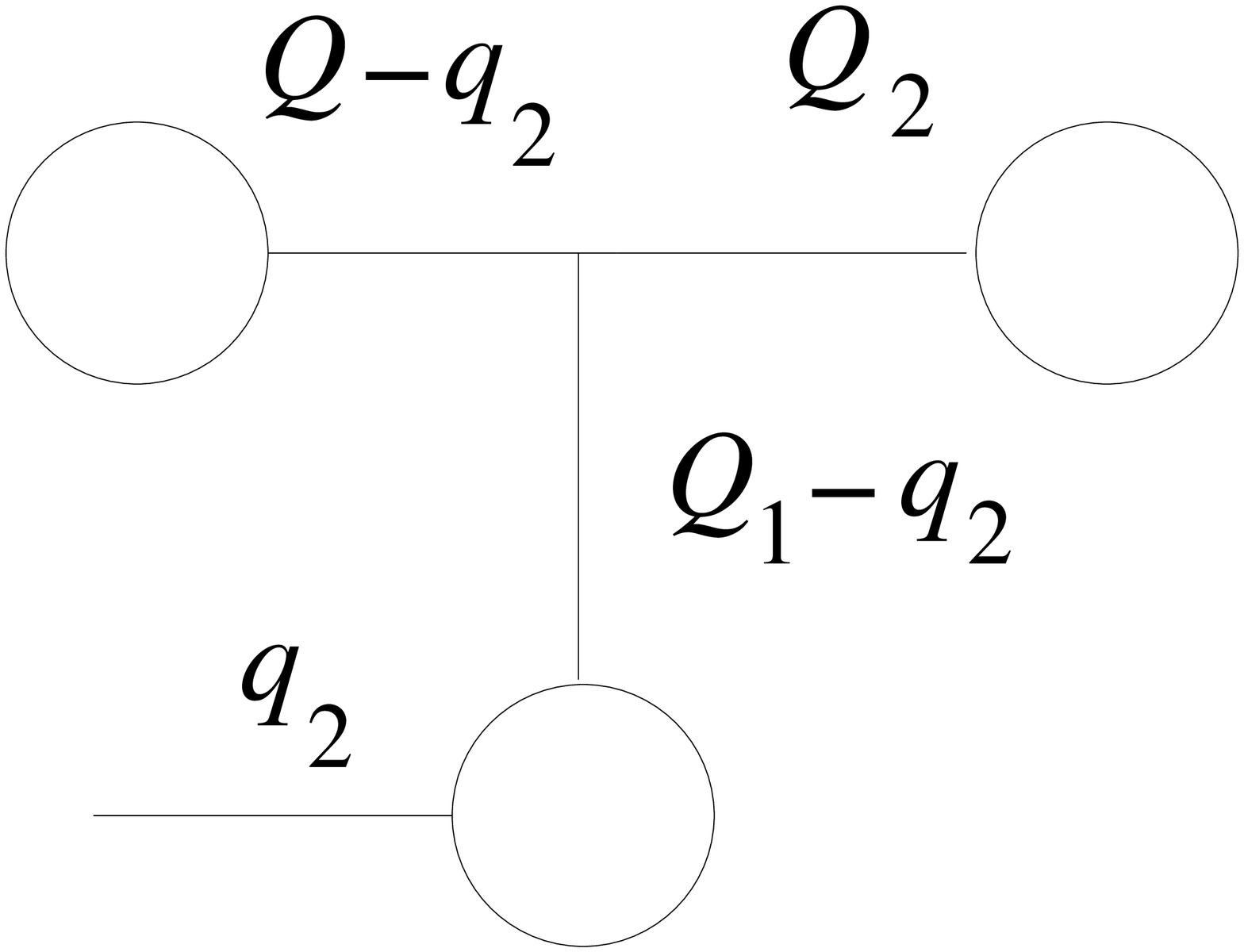}{3}{-24}
\]
\bqa
d\Phi_n &=& \ldots \frac{d Q_1^2}{2 \pi}\; \frac{d Q_2^2}{2 \pi}\;
d\Phi_2(Q \to Q_1,Q_2) \ldots \nl
&=&  \ldots \frac{d Q_1^2}{2 \pi} \;\frac{d Q_2^2}{2 \pi}\;
dt\; d\phi\; \frac{1}{32 \pi^2\; Q\; |\vec{q}_2|}\ldots
\nn\label{t-channel}
\eqa
with
\bqa t&=&(Q_1 - q_2 )^2
\nl
&=&m_2^2+Q_1^2-\frac{E_2}{Q}(Q^2+Q_1^2-Q_2^2)+
\frac{\lambda^{1/2}}{Q}|\vec{q}_2| \cos\theta 
\nn\eqa
and $(E_2,\vec{q}_2)$ being the incoming momentum $q_2$ in the
rest frame of $Q$.
The appropriate  sequence of vertices, $\{V_1, V_2, \ldots, V_k\}$ 
can be chosen in such a way that a recursive construction of the phase space
is realized. For instance $V_1$ should contain at least one incoming 
particle whose momentum is known. The rest of the sequence 
is chosen recursively: vertex $V_j$
is characterized by an incoming momentum $Q$ which has already been generated
in one of the $\{V_1,\ldots,V_{j-1}\}$.

Following the above described algorithm we end up with an expression
for the phase-space density,
\[
d\Phi_n \to \prod ds_i {\cal P}_i(s_i) \; \prod dt_j {\cal P}_j(t_j) \;
\prod d\phi_k \prod d\cos\theta_l
\]
where $s_i$ and $t_j$ refer to the kinematical invariants
entering the propagator factors of the graph and $\phi_k$ and $\cos\theta_l$
represent center-of-mass angles needed to complete the phase space
parametrization. It is now straightforward to generate
$s_i$ and $t_j$ with probability densities ${\cal P}_i(s_i)$ and 
${\cal P}_j(t_j)$ that are automatically 
chosen accordingly to the nature of the propagating
particle.

Results, demonstrating the ability of {\tt PHEGAS-HELAC} to deal with
multiparticle processes, are presented in 
table~\ref{tab1}~\cite{phegas,minilep2}. 

\section{Summary and Outlook}

{\tt PHEGAS-HELAC} offers a framework for high-energy
phenomenology. It provides all necessary and sufficient tools
for efficient, reliable and automatic computation of helicity amplitudes
and cross section. The Standard Model, including QCD, has been fully
incorporated.
Higher-order corrections are in principle tractable within the
framework of Dyson-Schwinger equations and work is in progress
in order to include electroweak corrections as described in 
reference~\cite{bbc}.
New physics interactions and models, icluding the Minimal Supersymmetric
Standard Model and the trilinear gauge couplins will be considered
in the near future.

\vspace{10mm}

\noindent{\bf Acknowledgments}

\noindent
C.G.P. would like to acknowledge Fermilab and Argonne National Laboratory
for their kind hospitality.

\bibliographystyle{aipproc}

\begin{thebibliography}{1}

\bibitem{four-fermion}
D.~Bardin {\it et al.},
``Event generators for W W physics,''
hep-ph/9709270 and references therein.


\bibitem{comphep}
E.~E.~Boos, M.~N.~Dubinin, V.~A.~Ilin, A.~E.~Pukhov and V.~I.~Savrin,
``CompHEP: Specialized package for automatic calculations of elementary particle decays and collisions,''
hep-ph/9503280.
\\
V.~A.~Ilin, D.~N.~Kovalenko and A.~E.~Pukhov,
Int.\ J.\ Mod.\ Phys.\  {\bf C7} (1996) 761
[hep-ph/9612479].

\bibitem{grace}
T.~Ishikawa, T.~Kaneko, K.~Kato, S.~Kawabata, Y.~Shimizu and H.~Tanaka
                  [MINAMI-TATEYA group Collaboration],
KEK-92-19.
\\
F.~Yuasa {\it et al.},
``Automatic computation of cross sections in HEP: Status of GRACE system,''
hep-ph/0007053.


\bibitem{madg}
T.~Stelzer and W.~F.~Long,
Comput.\ Phys.\ Commun.\  {\bf 81} (1994) 357
[hep-ph/9401258].

\bibitem{helac}
A.~Kanaki and C.~G.~Papadopoulos,
``HELAC: A package to compute electroweak helicity amplitudes,''
hep-ph/0002082.

\bibitem{phegas}
C.~G.~Papadopoulos,
``PHEGAS: A phase space generator for automatic cross-section  computation,''
hep-ph/0007335.

\bibitem{berends}
F.~A.~Berends and W.~T.~Giele,
Nucl.\ Phys.\  {\bf B306} (1988) 759.

\bibitem{camo}
F.~Caravaglios and M.~Moretti,
Phys.\ Lett.\  {\bf B358} (1995) 332
[hep-ph/9507237].

\bibitem{qcd}
P.~Draggiotis, R.~H.~Kleiss and C.~G.~Papadopoulos,
Phys.\ Lett.\  {\bf B439} (1998) 157
[hep-ph/9807207].

\bibitem{alphaqcd}
F.~Caravaglios, M.~L.~Mangano, M.~Moretti and R.~Pittau,
Nucl.\ Phys.\  {\bf B539} (1999) 215
[hep-ph/9807570].

\bibitem{mpcl}
David M.~Smith,
Transactions on Mathematical Software {\bf 17} (1991) 273- 283.
http://www.lmu.edu/acad/personal/fa\-culty/dmsmith2/FMLIB.html

\bibitem{nextc}
F.~A.~Berends, C.~G.~Papadopoulos and R.~Pittau,
``NEXTCALIBUR: A four-fermion generator for electron positron  collisions,''
hep-ph/0011031;
``Four-fermion production in electron positron collisions
with  NEXTCALIBUR,'' hep-ph/0002249.


\bibitem{multi}
R.~Kleiss and R.~Pittau,
Comput.\ Phys.\ Commun.\  {\bf 83} (1994) 141
[hep-ph/9405257].

\bibitem{Ohl:1999jn}
T.~Ohl,
Comput.\ Phys.\ Commun.\  {\bf 120} (1999) 13
[hep-ph/9806432].

\bibitem{sarge}
P.~D.~Draggiotis, A.~van Hameren and R.~Kleiss,
Phys.\ Lett.\  {\bf B483} (2000) 124
[hep-ph/0004047].


\bibitem{minilep2}
M.~W.~Grunewald {\it et al.},
``Four fermion production in electron positron collisions,''
hep-ph/0005309.


\bibitem{bbc}
W.~Beenakker, F.~A.~Berends and A.~P.~Chapovsky,
Nucl.\ Phys.\  {\bf B573} (2000) 503
[hep-ph/9909472].

\end{thebibliography}

\end{document}